%% file: ecir2027.tex
\documentclass[runningheads]{llncs}

\usepackage[T1]{fontenc}
\usepackage[utf8]{inputenc}
\usepackage{booktabs}
\usepackage{siunitx}
\usepackage{amsmath}
\usepackage{amssymb}
\usepackage{graphicx}
\usepackage{xcolor}
\usepackage[hidelinks]{hyperref}
\usepackage{fontawesome5}
\usepackage{cleveref}
\creflabelformat{equation}{#2\textup{#1}#3}
\usepackage{orcidlink}
\renewcommand{\orcidID}[1]{\unskip\,\orcidlink{#1}}

\newcommand{\ds}{ProdCrawl}

\begin{document}

\title{\ds: A Replayable Test Collection for Focused Crawling}
\titlerunning{\ds: A Replayable Test Collection for Focused Crawling}

\author{Manuel Rey-Escobar\inst{1,2}\orcidID{0009-0005-8955-509X} \and David Otero\inst{1}\orcidID{0000-0003-1139-0449} \and Javier Parapar\inst{1}\orcidID{0000-0002-5997-8252}}
\authorrunning{M. Rey-Escobar et al.}
\institute{IRLab, CITIC, Universidade da Coru\~na, A Coruña, Spain \and
Linknovate Science, Santiago de Compostela, Spain\\
\email{j.manuel.rey@udc.es}, \email{david.otero.freijeiro@udc.es}, 
\email{javier.parapar@udc.es}
}

\maketitle

\begin{abstract}
A focused crawler searches for web pages on a specific topic under a constrained download budget. Since it can only fetch a limited number of pages, its success depends on choosing which links to follow. Focused crawlers are usually evaluated on live websites by harvest rate, the fraction of downloaded pages that are relevant. This has two problems: on the one hand, websites change over time, so a crawl cannot be repeated under the same conditions; on the other hand, there is no ground truth for which pages are relevant. As a result, comparisons between crawlers are unreliable. We present \ds{}, a replayable test collection for focused crawling. It comprises static mirrors of 621 company and organization websites, about 1.1 million pages, with each page labeled as a product page or non-product page based on a majority vote among three large language models. Any crawler can be run on exactly the same sites and scored against the same labels. We use \ds{} to compare simple baselines (breadth-first, depth-first and text classifiers) with an approach that scores each link by searching for similar link texts whose target pages have already been labeled. Our results show that text classifiers and keyword matching do not significantly outperform breadth-first crawling, whereas neural embeddings, which compare link texts by semantic meaning, outperform all simple strategies. Because every run evaluates over identical pages and labels, these results are exactly reproducible. The collection, codebase, and models are publicly released to support experimental replication and fair crawler comparison.

\end{abstract}

\begin{center}
\href{https://github.com/IRLab-UDC/prodcrawl-experiments}{\faGithub\ \texttt{github.com/IRLab-UDC/prodcrawl}}
\end{center}

\input{ecir/introduction.tex}
\input{ecir/related_work.tex}
\input{ecir/collection.tex}
\input{ecir/method.tex}
\input{ecir/experiments.tex}
\input{ecir/conclusions.tex}

\bibliographystyle{splncs04}
\bibliography{references}

\end{document}

%% file: ecir/introduction.tex
\section{Introduction}
\label{sec:intro}

The term \emph{web crawling} designates the automated technologies used to systematically and efficiently discover and gather content from the Web~\cite{olston2010web}. A crawler initiates traversal from one or more designated seed URLs, and systematically explores newly discovered hyperlinks by following predefined crawling policies. While generic crawlers aim to index entire web segments, \emph{focused crawlers} perform targeted vertical explorations to retrieve pages relevant to a specific topic~\cite{chakrabarti1999focused}. Because crawlers operate under constrained resource limits, a focused crawler must optimize its download budget (the maximum volume of pages permitted across target domains). The critical challenge lies in link prioritization: ordering unvisited candidate URLs on the frontier so as to discover relevant pages early while avoiding irrelevant paths~\cite{cho1998efficient}.

The standard performance measure of a focused crawler is harvest rate~\cite{chakrabarti1999focused,batsakis2009improving,pant2005learning}, the proportion of downloaded pages that are relevant. It is typically measured by running the crawler on the live Web, starting from real seed URLs and judging the pages it downloads. This practice, however, presents three fundamental challenges. First, computing harvest rate requires determining whether every fetched page is relevant, yet the live Web provides no standardized ground truth. Chakrabarti et al.\ originally evaluated their crawler using the same classifier that guided its link traversal~\cite{chakrabarti1999focused}, a circular dependency that subsequent work continues to replicate~\cite{kuehn2023threatcrawl}. Also, because the complete set of relevant pages across target domains remains unknown, recall cannot be reliably measured~\cite{pant2005learning}. Second, live Web experiments are vulnerable to external variables beyond the experimenter's control, including server outages, request throttling, asynchronous network latency, and continuous content updates on target sites~\cite{meusel2014focused,klein2018focused,pezzuti2025neural}. Thus, replicating a live crawl under identical operating conditions is virtually impossible. Third, crawlers with distinct ranking policies traverse divergent paths and encounter different content. Because live web content continuously drifts, harvest rates reported across different studies, or even across successive runs of the same study, cannot be directly compared. Isolating whether an effectiveness gain stems from an improved link-ranking algorithm or from arbitrary shifts in the underlying web graph remains impossible.

To overcome these challenges, we introduce \ds{}\footnote{https://huggingface.co/datasets/irlab-udc/prodcrawl}, a replayable test collection for focused crawling. \ds{} provides complete static mirrors of \num{621} company and organization websites, paired with ground-truth product-page labels for over \num{1.1}~million pages across these domains. By evaluating every crawler against identical, pre-computed labels rather than an internal classifier, \ds{} establishes an objective ground truth and enables the measurement of recall alongside harvest rate. The static mirrors ensure experimental control: crawlers encounter identical responses on every execution, unaffected by server latency or remote content drift. Together, the mirrored environments and objective labels guarantee rigorous comparability across different algorithms. Alongside the collection, we release a reproduction codebase and our task-adapted bi-encoder model\footnote{https://huggingface.co/irlab-udc/anchor-product-page-mpnet-base}.

Our retrieval-based anchor scoring approach addresses link prioritization as an information retrieval task over anchor-text evidence stores, evaluating lexical (BM25~\cite{robertson2009probabilistic}), dense ($k\text{NN}$ and $a\text{NN}$~\cite{malkov2020efficient}), hybrid, and task-adapted bi-encoder representations~\cite{reimers2019sentence-bert}. By decoupling individual ranking components, our framework enables modular, single-variable benchmarking of crawl strategies under strictly controlled, identical conditions.

%% file: ecir/related_work.tex
\section{Related Work}
\label{sec:rw}

The core mechanism of a focused crawler is link prioritization: determining which candidate URLs on the unvisited frontier to explore next under a constrained download budget~\cite{chakrabarti1999focused,cho1998efficient}. Early approaches guided traversal using anchor keyword matching~\cite{cho1998efficient} or Web graph topology~\cite{brin1998anatomy,menczer2001links}. Chakrabarti et al.\ introduced statistical page classification for frontier scoring~\cite{chakrabarti1999focused} and online adaptation during traversal~\cite{chakrabarti2002accelerated}. Subsequent works expanded this paradigm by training over harvested text~\cite{barbosa2007adaptive,du2015improved}, context graphs~\cite{guan2016design,hsu2006topic-specific}, or topic ontologies~\cite{nayak2012discovering,wang2017simulation}, while others framed selection as a search heuristic using genetic algorithms~\cite{wang2016research} or ant colony optimization~\cite{chen2011searching} under exploration-exploitation trade-offs~\cite{meusel2014focused,pant2005learning}. Crucially, these strategies rely on fixed parametric models requiring expensive retraining to incorporate new signals. In contrast, retrieval-based anchor scoring compares candidates directly against instance-level anchor evidence, enabling dynamic index updates via simple insertion.

Neural representation learning has shifted link scoring from exact lexical matching toward dense semantic alignment. While lexical overlap misses paraphrased terminology and cross-encoders are too slow for real-time link scoring, bi-encoders like Sentence-\hspace{0pt}BERT~\cite{reimers2019sentence-bert} enable low-latency dense retrieval for crawling. This architecture has driven adoption across domain-specific crawlers \cite{kuehn2023threatcrawl,rajiv2023supervised,tawil2021bert}: Rajiv and Navaneethan~\cite{rajiv2023supervised} used GloVe embeddings without iterative frontier prioritization; Tawil and Alqaraleh~\cite{tawil2021bert} categorized pages via Sentence-\hspace{0pt}BERT; and Kuehn et al.\ developed \emph{Threat\-Crawl}~\cite{kuehn2023threatcrawl} to evaluate candidates against static representations~\cite{koloveas2021intime}. Recently, pretrained language models have also been applied to general collection, ranking candidates by estimated document quality~\cite{pezzuti2025document,pezzuti2025neural} or LLM pre-training utility~\cite{yu2025craw4llm}.

Despite progress from heuristics to neural rankers, empirical crawler evaluation remains a major bottleneck. Early benchmarks evaluated against live directories like ODP/\slash DMOZ or the live Web~\cite{chakrabarti1999focused,srinivasan2005general}, but live evaluations suffer from content drift and network volatility~\cite{srinivasan2005general}. Researchers thus turned to frozen web corpora (e.g., WebKB~\cite{craven1998learning}, WT10g~\cite{bailey2003engineering}, GOV2~\cite{clarke2004overview}, ClueWeb~\cite{callan2009clueweb09,overwijk2022clueweb22}, Common Crawl subgraphs~\cite{crawlfoundation2024common,meusel2014focused}, and ORCA~\cite{rder2021orca}) for simulated crawls~\cite{pezzuti2025document,pezzuti2025neural,yu2025craw4llm}. Critically, however, these collections lack task-specific crawling judgments, providing only ad-hoc retrieval annotations or broad document categories. Conversely, while product datasets support entity matching~\cite{peeters2024wdc} and search~\cite{reddy2022shopping}, they strip away the hyperlink graph topology necessary for navigation.

%% file: ecir/collection.tex
\section{The \ds{} Collection}
\label{sec:collection}

To establish a reproducible benchmark for focused crawling, \ds{} targets the discovery of \emph{product pages} across corporate and organizational websites. Product pages represent high-value transactional endpoints—pages specifically designed to facilitate conversion on an offering (such as a physical item, software license, or service tier). Because product pages constitute a sparse minority on most organizational sites, a crawler's link prioritization strategy directly determines whether its download budget reaches them before exhausting resources.

Benchmarking focused crawlers on this task requires two parts:
\begin{enumerate}
    \item \textbf{Navigable web}: Static, hyperlinked website mirrors that allow crawlers to explore realistic graph structures and replay identical download requests deterministically.
    \item \textbf{Labels}: Objective, pre-computed relevance judgments for every reachable page to evaluate precision and recall without circular classifier dependencies.
\end{enumerate}
\subsection{Site Acquisition}
\label{sec:acquisition}

\noindent\textbf{Site Selection.} 
To identify target domains, we queried Linknovate's aggregated indices\footnote{\url{https://www.linknovate.com/}} for company and organization websites. From the returned candidates, we filtered domains to retain \num{621} sites that contained at least one verified product page, ensuring a meaningful target density for focused evaluation. The resulting corpus is predominantly in English, which serves as the primary language for \num{95}\% of the included sites.

\noindent\textbf{Static Web Mirroring.} 
To eliminate temporal content drift and server-side variability, we archived each web domain using the HTTrack website copier~\cite{roche2023httrack}, preserving its complete HTML content, hyperlink graph, and directory hierarchy. Crawlers evaluated on \ds{} issue standard HTTP requests against a local host of these mirrors, ensuring byte-identical responses across independent runs.

\subsection{Relevance Annotation and Quality}
\label{sec:annotation}

\noindent\textbf{Relevance Definition.} 
Following the current use of instruction-tuned LLMs as relevance assessors in Information Retrieval~\cite{faggioli2023perspectives,thomas2024large}, every page in the mirror was classified according to its plain text.  

A page is labeled as a \emph{product page} (relevant) if its primary purpose is bottom-of-the-funnel conversion for a specific offering (e.g.,\ an SKU, software product, or tiered service plan), indicated by pricing tables, add-to-cart controls, feature specs, or demo request forms. Conversely, non-product pages comprise top- and middle-of-the-funnel content like news, documentation, landing pages, and corporate overviews. Under a strict tie-breaking rule, any page whose body is over half narrative or educational (e.g.,\ guides or blogs) is classified as non-product, even if it links to commercial products.

\noindent\textbf{LLM Jury.} 
To mitigate individual model bias, each extracted text was independently evaluated by three LLMs: Llama~3.1 8B Instruct~\cite{grattafiori2024llama}, Llama~4 Scout 17B-16E Instruct~\cite{ai2025llama}, and Gemma~3 27B IT~\cite{team2025gemma}. Each model was presented with an identical system instruction framing the task as a binary classification decision and was constrained to output either \emph{Yes} or \emph{No} via greedy decoding in vLLM~\cite{kwon2023efficient}. The final ground-truth label for each page is assigned by majority vote.

\input{ecir/tables/tab_agreement.tex}

\noindent\textbf{Annotator Agreement.} 
\Cref{tab:agreement} reports the inter-model agreement across all splits. Fleiss' $\kappa$~\cite{fleiss1971measuring} ranges from \num{0.76} to \num{0.85}, denoting substantial to near-perfect consensus among the three annotators~\cite{landis1977measurement}. Furthermore, between \num{84}\% and \num{89}\% of all pages receive a unanimous vote from all three models. Disagreement is moderately higher on the positive class, this does not undermine benchmark validity: a link whose target page receives two out of three affirmative votes remains an substantially more relevant crawl target than one rejected unanimously, and majority voting effectively captures this relative crawling priority.

\subsection{Collection Statistics}
\label{sec:splits}

The complete dataset contains \num{1102934} web pages across \num{621} sites, of which \num{397451} (\num{36.0}\%) are labeled as relevant product pages. We partition the dataset into disjoint training, validation, and test collections. The main statistics of the dataset are presented in \Cref{tab:composition}.
\input{ecir/tables/tab_composition.tex}
The partition is performed strictly by site. We allocate sites randomly, assigning approximately twice as many sites to training as to the evaluation splits

%% file: ecir/tables/tab_agreement.tex
\begin{table}[t]
\centering
\caption{Agreement of the three labeling models over the pages of each split. The
page label is the majority of the three votes. \emph{Unanimous} pages received
the same vote from all three models; the last two rows give that share among the
pages labeled relevant and among the others.}
\label{tab:agreement}
\footnotesize
\setlength{\tabcolsep}{9pt}
\begin{tabular}{lrrr}
\toprule
& Train & Validation & Test \\
\midrule
\multicolumn{4}{l}{\emph{Pages labeled relevant (\%)}} \\
\quad Llama 3.1 & 38.1 & 21.2 & 45.4 \\
\quad Llama 4 Scout & 36.6 & 21.3 & 33.7 \\
\quad Gemma 3 & 43.4 & 28.3 & 47.4 \\
\quad Majority & 38.3 & 24.2 & 45.2 \\
\midrule
\multicolumn{4}{l}{\emph{Cohen's $\kappa$}} \\
\quad Llama 3.1 / Gemma 3 & .82 & .77 & .92 \\
\quad Llama 3.1 / Llama 4 Scout & .92 & .75 & .72 \\
\quad Gemma 3 / Llama 4 Scout & .82 & .78 & .71 \\
\midrule
Fleiss' $\kappa$ & .85 & .76 & .78 \\
Unanimous pages (\%) & 89.4 & 87.3 & 84.0 \\
\quad of relevant pages (\%) & 90.4 & 70.2 & 72.0 \\
\quad of other pages (\%) & 88.7 & 92.7 & 93.8 \\
\bottomrule
\end{tabular}
\end{table}

%% file: ecir/tables/tab_composition.tex
\begin{table}[t]
\centering
\caption{Composition of the three collections. Per-site values are means over
sites; the percentage is the mean share of relevant pages per site. Depth is the
mean shortest link distance from the seed page to a relevant page.}
\label{tab:composition}
\footnotesize
\setlength{\tabcolsep}{5pt}
\begin{tabular}{lrrrrrr}
\toprule
& & & & \multicolumn{2}{c}{Per site} & \\
\cmidrule(lr){5-6}
Split & Sites & Pages & Relevant (\%) & Pages & Relevant (\%) & Depth \\
\midrule
Train & \num{318} & \num{544059} & \num{208124} (38.3) & \num{1711} & \num{654} (24.0) & 3.99 \\
Validation & \num{167} & \num{301312} & \num{72847} (24.2) & \num{1804} & \num{436} (24.7) & 3.12 \\
Test & \num{136} & \num{257563} & \num{116480} (45.2) & \num{1894} & \num{856} (20.4) & 3.49 \\
\midrule
All & \num{621} & \num{1102934} & \num{397451} (36.0) & \num{1776} & \num{640} (23.4) & 3.68 \\
\bottomrule
\end{tabular}
\end{table}

%% file: ecir/method.tex
\section{Prioritization Methods}
\label{sec:method}

A focused crawler iteratively dequeues URLs from an unvisited frontier, downloads their content, extracts outbound hyperlinks, and assigns each candidate link a priority before enqueuing it. Built on top of the Scrapy framework, our crawler orchestrates traversal through a centralized priority queue. 

Candidate links are prioritized by a scoring function $s(a) \in [0, 1]$ evaluated on the discovering link's anchor text $a$, where higher scores represent higher predicted utility. When two links share identical priority scores, the frontier breaks ties deterministically using the discovery sequence order $\tau(a)$, preserving FIFO ordering among equally scored candidates. Below, we formalize the classical baseline policies and our proposed retrieval-based framework.

\subsection{Classical Crawling Baselines}
\label{sec:baselines}

\noindent\textbf{Uninformed Baselines.} 
Breadth-first search (BFS) and depth-first search (DFS) ignore anchor text and rely solely on graph topology. BFS prioritizes links strictly in discovery order $\tau(a)$ (FIFO), ensuring an exhaustive, shallow exploration of the domain. DFS inverts this order (LIFO), aggressively following deep navigational chains.

\noindent\textbf{Supervised Classifiers.} 
Following the predominant paradigm in classical focused crawling~\cite{chakrabarti1999focused,du2015improved}, we evaluate linear Support Vector Machines (SVM) and Multinomial Naive Bayes (NB) classifiers. Both models are trained on TF--IDF unigram vectors extracted from anchor texts in the training collection. The classifiers output binary predictions $\hat{y} \in \{0, 1\}$, partitioning the frontier into predicted relevant ($s(a) = 1$) and irrelevant ($s(a) = 0$) candidate pools, with discovery order $\tau(a)$ governing traversal within each partition.

\subsection{Retrieval-Based Anchor Scoring}
\label{sec:scoring}

Rather than training a rigid parametric classifier that requires full retraining to incorporate new data, \emph{Retrieval-Based Anchor Scoring} (RBAS) casts link prioritization as an information retrieval query against an indexed evidence store $\mathcal{E}$ of previously labeled anchor texts.

\subsubsection{Scoring Formulation.}

Let $\mathcal{E}$ denote an evidence store of previously observed anchor texts, where each entry $e \in \mathcal{E}$ is associated with the ground-truth relevance label of its target page, denoted by the indicator $\mathrm{rel}(e) \in \{0, 1\}$. When the crawler encounters an unvisited candidate link, it treats its anchor text $a$ as a query issued against $\mathcal{E}$. 

A chosen retrieval model computes a similarity score $\sigma(a, e)$ quantifying the affinity between candidate anchor $a$ and any stored anchor $e \in \mathcal{E}$. The retrieval backend then returns the top-$k$ entries in $\mathcal{E}$ with the highest scores, forming the retrieved neighborhood $N_k(a) \subset \mathcal{E}$. Finally, the candidate link's priority score $s(a) \in [0, 1]$ is defined as the similarity-weighted proportion of retrieved score mass that belongs to relevant targets:
\begin{equation}
  s(a) \;=\;
  \frac{\sum_{e \in N_k(a)} \mathbf{1}[\mathrm{rel}(e)]\;\sigma(a,e)}
       {\sum_{e \in N_k(a)} \sigma(a,e)} ,
  \label{eq:score}
\end{equation}
where $\mathbf{1}[\cdot]$ is the indicator function, and $s(a) = 0$ if no documents are retrieved ($N_k(a) = \emptyset$).

Under this formulation, $s(a)$ acts as a non-parametric, similarity-weighted $k$-nearest-neighbor estimator of the probability that following link $a$ will lead to a relevant page. Crucially, \Cref{eq:score} is invariant to any positive linear rescaling of $\sigma$ (i.e., replacing $\sigma$ with $c \cdot \sigma$ for any constant $c > 0$). This invariance ensures that sparse lexical, dense neural, and hybrid scoring functions—whose raw outputs inhabit fundamentally different numerical ranges—can be evaluated on an identical, normalized scale without requiring external calibration.

\subsubsection{Retrieval Backends.}
We instantiate the scoring function $\sigma(a, e)$ across four distinct retrieval architectures spanning modern IR:
\begin{itemize}
    \item \textbf{Lexical (BM25)}: Scores query anchors against an inverted index over target anchor texts using standard BM25~\cite{robertson2009probabilistic} ($k_1 = 1.2$, $b = 0.75$).
    \item \textbf{Dense Exact ($k\text{NN}$)}: Encodes candidate anchors into dense vector representations using a bi-encoder and exhaustively calculates the cosine similarity $\sigma(a, e) = 1 + \cos(a, e)$ over all entries in $\mathcal{E}$.
    \item \textbf{Dense Approximate ($a\text{NN}$)}: Replaces exhaustive vector comparison with approximate nearest neighbor search over a Hierarchical Navigable Small World (HNSW) graph~\cite{malkov2020efficient}, evaluating $1000$ candidates per query to bound decision latency.
    \item \textbf{Hybrid (BM25$\rightarrow k\text{NN}$)}: A two-stage ranker where BM25 first retrieves a candidate window of $w$ entries sharing lexical terms with $a$, and the bi-encoder subsequently rescores and ranks them by cosine similarity, returning the top-$k$.
\end{itemize}

\subsubsection{Evidence Stores.}
We evaluate three evidence store configurations to study cross-domain and online knowledge transfer:
\begin{itemize}
    \item \textbf{Offline Store ($\mathcal{E}_{\mathrm{off}}$)}: A static index holding the labeled anchor texts from the disjoint training collection, representing prior knowledge gathered from external domains.
    \item \textbf{Online Store ($\mathcal{E}_{\mathrm{on}}$)}: A dynamic index that initializes empty and is populated on-the-fly with downloaded pages and their verified ground-truth labels during traversal, representing adaptive domain exploration.
    \item \textbf{Combined Store ($\mathcal{E}_{\mathrm{comb}}$)}: Evaluates queries simultaneously across both offline and online indices ($\mathcal{E}_{\mathrm{off}} \cup \mathcal{E}_{\mathrm{on}}$).
\end{itemize}

\subsubsection{Task-Specific Encoder Adaptation.}
\label{sec:adaptation}
Under dense RBAS, \Cref{eq:score} is only effective if the embedding space cleanly clusters anchors of relevant target pages together while separating them from irrelevant ones. Off-the-shelf bi-encoders such as \texttt{all-mpnet-base-v2}~\cite{song2020mpnet,reimers2019sentence-bert} are optimized for general semantic similarity rather than identifying transactional product intent. To align representations with this vertical search task, we fine-tune the bi-encoder on the training collection. Supervision is defined over normalized anchor text types rather than raw link occurrences. From \num{200000} labeled training links, we determine the empirical relevance ratio $\rho(t)$ for each distinct anchor string $t$. We retain unambiguous types with $\rho(t) \ge 0.9$ as reliably relevant and $\rho(t) \le 0.1$ as reliably irrelevant, discarding ambiguous intermediate anchors and uninformative generic navigation tokens (e.g.,\ \emph{read more}, \emph{previous}, or bare numbers). 
To structure the learned metric space, we anchor the encoder around $15$ domain-specific reference phrases (such as structural descriptors like "our products" and "our service") rather than arbitrary query keywords. We train the model via cosine similarity regression over scored pairs: prototype-to-relevant anchor ($y = 1.0$), closely related anchors ($y = 0.9$), and relevant versus irrelevant anchors ($y = 0.0$). Fine-tuning runs for one epoch with a batch size of $32$ and a learning rate of $2 \times 10^{-5}$. At crawl time, candidate texts undergo identical normalization and noise filtering prior to retrieval, tightly coupling encoder adaptation with domain-specific query sanitization.

%% file: ecir/experiments.tex
\section{Experiments}
\label{sec:results}

We explore the collection by running each prioritization approach of
\Cref{sec:method} on it. \Cref{sec:setup} describes the setup. \Cref{sec:exp-results}
reports each approach against the baselines.

\subsection{Setup}
\label{sec:setup}

\subsubsection{Crawl simulation.}
The crawler is built on Scrapy, and each evidence store is an Elasticsearch~8.12 index. 
Each run crawls every site of a collection against an isolated local mirror, using 
one crawler process per site (restricted to its domain with one request in flight) and 
crawling up to 50 sites concurrently. To evaluate performance up to a budget of 
$b = 301$ in validation and $b = 257$  in test downloads—representing approximately 10\% of the mean number of pages per site—crawling 
continues well beyond this threshold. From each downloaded page, 
the crawler scores only the first 100 internal links in document order. This bounds 
the number of retrieval queries per downloaded page, analogous to the default outlink 
limit in Apache Nutch~\cite{apachesoftwarefoundation2024apache}. 

\subsubsection{Measures.}
Let $D$ be the sites of a collection, $R_d$ the number of relevant pages that
site $d$ holds, and $p^d_1, \dots, p^d_b$ the first $b$ pages the crawler
downloaded from $d$. We report mean average precision at crawl budget $b$,
\begin{equation}
  \mathrm{MAP}@b \;=\; \frac{1}{|D|}\sum_{d \in D}
  \frac{1}{R_d}\sum_{\substack{i \le b\\ \mathrm{rel}(p^d_i)}}
  \frac{\bigl|\{\, j \le i : \mathrm{rel}(p^d_j) \,\}\bigr|}{i} .
  \label{eq:map}
\end{equation}
Precision at step $i$ is the harvest rate at that
step~\cite{chakrabarti1999focused}. \Cref{eq:map} therefore aggregates harvest
rate over the crawl and rewards early downloads. Normalizing by $R_d$ makes sites
with very different numbers of relevant pages comparable. Absolute values are well
below one because $b \ll R_d$ on large sites; \Cref{tab:composition} gives
the attainable ceiling. We evaluate budgets up to $b = 30$ and also report
$\mathrm{AUC}$, the trapezoidal area under $\mathrm{MAP}@b$ for $b \in [1, 30]$.
Sites whose reachable pages run out before 30 downloads (18 on validation, 16 on
test) are truncated identically for every system. Significance is a two-sided
Wilcoxon signed-rank test over per-site average precision at $b = 30$, paired by
site.

\subsubsection{Hyperparameter Tuning and Model Selection.}
\label{sec:grid-selection}
\begin{figure}[t]
\centering
\includegraphics[width=0.86\textwidth]{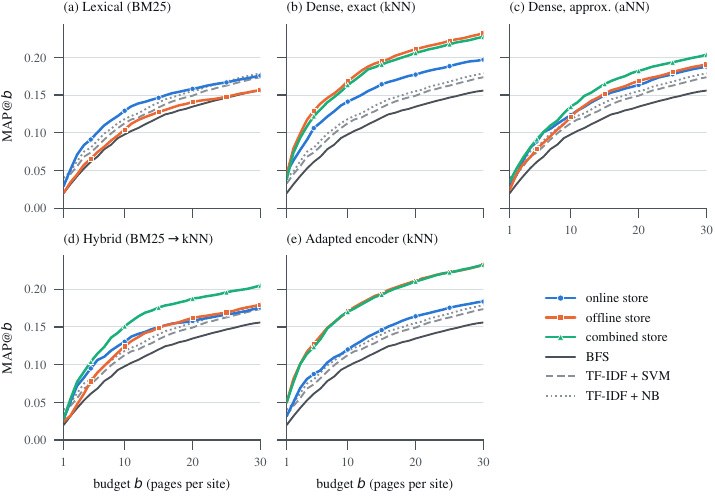}
\caption{$\mathrm{MAP}@b$ on the test collection (136 sites) for the
configurations of \Cref{tab:reference}, one panel per experiment and one line
per evidence store. Panels (a) to (d) use the generic encoder; (e) shows kNN with the
adapted encoder. Grey lines are baselines; DFS, far below all policies, is omitted.}
\label{fig:results}
\end{figure}

We optimize model hyperparameters on the 167 validation sites via grid search, evaluating 218 total configurations alongside four baselines. We vary the parameter $k \in \{5, 10, 15, 20, 50\}$  across all retrieval paradigms. Additionally, on the hybrid (BM25 $\rightarrow$ Dense): we swept $k$ values for every window sizes $w \in \{50, 100, 150, 200, 500\}$ (excluding $k = w = 50$ as it makes no sense).
The baselines were executed under identical crawling conditions. Because baseline priorities (BFS discovery counters, DFS counters, or binary classifier decisions) operate independently of our priority resolution (\Cref{sec:method}), their execution order remains invariant.
For each of the 20 core model configurations (combining retrieval model, store, and encoder), we select the hyperparameter setup achieving the highest AUC ($b=30$) on validation. These selected models are evaluated directly on the test set without re-tuning (\Cref{tab:reference}, \Cref{fig:results}).

\input{ecir/tables/tab_reference.tex}

\subsection{Results}
\label{sec:exp-results}

\subsubsection{Lexical vs.\ Dense Matching Performance.}
\label{sec:exp-lexical-dense}
In the validation split, we begin by evaluating the baseline retrieval paradigm against dense matching. Scoring links using BM25 over the online and offline stores fails to yield significant improvements over uninformed crawling; neither configuration significantly outperforms BFS on either split ($p \ge 0.10$, with the best variant reaching $1.15\times$ MAP@30 on validation and $1.13\times$ on test). Similarly, neither classifier significantly beats BFS ($p \ge 0.06$), indicating that labeled evidence from other sites does not improve term matching. 

Conversely, replacing exact term matching with dense representations via exact kNN search (using the generic encoder) yields substantial gains, outperforming BM25 ($+.031$ MAP@30, $p = 0.008$). The performance boost is particularly pronounced when using the offline store ($+.053$) compared to the online store alone ($+.010$). Across all dense configurations, the evidence store proves crucial: the offline store outperforms the online store in 53 of 68 validation configurations ($+.017$, $p < 10^{-6}$), while the combined store achieves the highest performance, outperforming the offline store in 54 of 69 cases ($+.013$, $p < 10^{-9}$).

\subsubsection{Impact of Approximate Search (Exact vs.\ HNSW).}
\label{sec:exp-approximate}
To assess the trade-off between retrieval latency and crawling effectiveness, we replace exact kNN with HNSW approximate search (aNN). Exact search maintains a statistically significant advantage over approximate search across both the offline store ($+.037$, $p = 0.002$) and the combined store ($+.017$, $p = 0.002$), indicating that approximation errors become noticeable as the index scale grows. Over the smaller online store, however, the difference between exact and approximate search is negligible ($-.006$, $p = 0.064$). On the test set, exact search consistently leads in all six selected pairs (Fig.~\ref{fig:results}c).

\subsubsection{Effectiveness of Hybrid Ranking.}
\label{sec:exp-cascade}
We next evaluate whether a hybrid pipeline, i.e. using a lexical BM25 first stage to feed dense rescoring, improves upon single-stage dense retrieval. Averaged across window sizes $w$, the BM25$\rightarrow$kNN hybrid exceeds exact kNN in only 9 of 30 matched pairs ($-.008$, $p = 0.011$), though it outperforms standalone aNN in 21 of 30 pairs ($+.008$, $p = 0.012$). Rather than enhancing peak effectiveness, the hybrid acts as a computationally cheaper approximation of exact search. Varying the window size $w$ yields minimal impact on overall effectiveness (mean MAP@30 remaining between $.196$ and $.198$). With the adapted encoder over the combined store, the hybrid achieves the highest MAP@30 on validation and ranks fourth by AUC on test, performing comparably to the top system ($p = 0.40$; \Cref{fig:results}).

\subsubsection{Contribution of Domain-Specific Encoder Adaptation.}
\label{sec:exp-adaptation}
Having established dense retrieval as the superior paradigm, we evaluate the impact of our domain-specific encoder adaptation (\Cref{sec:adaptation}). Integrating the adapted encoder into kNN, aNN, and hybrid models drives consistent performance gains over the offline store, improving MAP@30 in 31 of 34 of the validation configurations ($+.023$, $p < 10^{-7}$). Over the combined store, improvements remain statistically significant ($+.008$, $p < 10^{-3}$). Conversely, performance over the online store slightly decreases ($-.003$, $p = 0.005$); a plausible cause is the generic-anchor filter removing evidence from a store that is still nearly empty. On test, the adapted encoder leads in all six offline and combined paired configurations, with kNN over the offline store reaching the highest overall AUC (\Cref{fig:results}e).

\subsubsection{Generalization to Held-Out Test Sites.}
\label{sec:transfer}
Finally, we evaluate the cross-collection transfer of the 24 fixed configurations from validation to test without any retuning. The overall ranking of systems demonstrates high stability, with validation and test AUC values correlating strongly (Spearman $\rho = 0.90$, $p < 10^{-8}$). On the test collection, the top configuration selected from validation achieves $1.45\times$ the MAP@30 of BFS and $1.31\times$ that of the SVM baseline. In total, 10 of the 20 configurations significantly outperform all four baselines on test—all leveraging either the offline or combined evidence store.

%% file: ecir/tables/tab_reference.tex
\begin{table}[tp]
\centering
\small
\caption{Reference results on the validation (167 sites) and test
(136 sites) collections. Each retrieval row is the configuration of
retrieval model, evidence store and encoder with the highest AUC on validation
($k$ neighbors, re-ranking window $w$), run once on test. $^{\textsc{a}}$~marks the
adapted encoder (Exp.~5). Bold is the best value in a column; $^\dagger$ marks a
$\mathrm{MAP}@30$ significantly higher than all four baselines on that
collection (two-sided Wilcoxon signed-rank test over sites, $p < 0.05$).}
\label{tab:reference}
\footnotesize
\setlength{\tabcolsep}{4pt}
\renewcommand{\arraystretch}{0.92}
\begin{tabular}{ll r@{}l r@{}l r@{}l r@{}l}
\toprule
& & \multicolumn{4}{c}{Validation} & \multicolumn{4}{c}{Test} \\
\cmidrule(lr){3-6}\cmidrule(lr){7-10}
Store & Config. & \multicolumn{2}{c}{MAP@30} & \multicolumn{2}{c}{AUC} & \multicolumn{2}{c}{MAP@30} & \multicolumn{2}{c}{AUC} \\
\midrule
\multicolumn{10}{l}{\emph{No retrieval}} \\
\quad BFS & -- &  & .1579 & 3 & .394 &  & .1560 & 3 & .218 \\
\quad DFS & -- &  & .0841 & 1 & .572 &  & .0669 & 1 & .218 \\
\quad TF--IDF + SVM & -- &  & .1694 & 3 & .758 &  & .1736 & 3 & .646 \\
\quad TF--IDF + NB & -- &  & .1699 & 3 & .741 &  & .1788 & 3 & .815 \\
\midrule
\multicolumn{10}{l}{\emph{Lexical retrieval: BM25 (Exp.~1)}} \\
\quad online & $k{=}20$ &  & .1816 & 4 & .046 &  & .1758 & 3 & .959 \\
\quad offline & $k{=}5$ &  & .1656 & 3 & .547 &  & .1567 & 3 & .353 \\
\midrule
\multicolumn{10}{l}{\emph{Dense retrieval, exact: kNN (Exp.~2)}} \\
\quad online & $k{=}15$ &  & .1902 & 4 & .165 &  & .1970 & 4 & .432 \\
\quad offline & $k{=}10$ &  & .2237$^\dagger$ & 5 & .046 &  & .2321$^\dagger$ & 5 & .277 \\
\quad combined & $k{=}10$ &  & .2203$^\dagger$ & 4 & .940 &  & .2271$^\dagger$ & 5 & .127 \\
\quad online$^{\textsc{a}}$ & $k{=}20$ &  & .1963 & 4 & .227 &  & .1837 & 3 & .978 \\
\quad offline$^{\textsc{a}}$ & $k{=}10$ &  & .2211$^\dagger$ & 5 & .000 &  & .2321$^\dagger$ & \textbf{5} & \textbf{.301} \\
\quad combined$^{\textsc{a}}$ & $k{=}10$ &  & .2207$^\dagger$ & 4 & .994 &  & \textbf{.2324}$^\dagger$ & 5 & .267 \\
\midrule
\multicolumn{10}{l}{\emph{Dense retrieval, approximate: aNN (Exp.~3)}} \\
\quad online & $k{=}20$ &  & .2068$^\dagger$ & 4 & .662 &  & .1877 & 4 & .030 \\
\quad offline & $k{=}25$ &  & .1840 & 3 & .934 &  & .1908 & 4 & .016 \\
\quad combined & $k{=}25$ &  & .2021$^\dagger$ & 4 & .513 &  & .2036$^\dagger$ & 4 & .382 \\
\quad online$^{\textsc{a}}$ & $k{=}20$ &  & .1908 & 4 & .158 &  & .1749 & 3 & .770 \\
\quad offline$^{\textsc{a}}$ & $k{=}10$ &  & .2045$^\dagger$ & 4 & .535 &  & .2171$^\dagger$ & 4 & .900 \\
\quad combined$^{\textsc{a}}$ & $k{=}10$ &  & .2117$^\dagger$ & 4 & .710 &  & .2318$^\dagger$ & 5 & .182 \\
\midrule
\multicolumn{10}{l}{\emph{Hybrid: BM25$\rightarrow$kNN (Exp.~4)}} \\
\quad online & $k{=}15$, $w{=}150$ &  & .1933 & 4 & .381 &  & .1754 & 3 & .986 \\
\quad offline & $k{=}10$, $w{=}100$ &  & .1907 & 4 & .117 &  & .1795 & 3 & .875 \\
\quad combined & $k{=}20$, $w{=}150$ &  & .2245$^\dagger$ & 5 & .057 &  & .2049$^\dagger$ & 4 & .621 \\
\quad online$^{\textsc{a}}$ & $k{=}5$, $w{=}150$ &  & .1859 & 4 & .199 &  & .1999 & 4 & .518 \\
\quad offline$^{\textsc{a}}$ & $k{=}5$, $w{=}50$ &  & .2233$^\dagger$ & 5 & .245 &  & .2235$^\dagger$ & 5 & .085 \\
\quad combined$^{\textsc{a}}$ & $k{=}10$, $w{=}50$ &  & \textbf{.2344}$^\dagger$ & \textbf{5} & \textbf{.470} &  & .2268$^\dagger$ & 5 & .260 \\
\bottomrule
\end{tabular}
\end{table}

%% file: ecir/conclusions.tex
\section{Conclusions}
\label{sec:conclusions}

Focused crawlers are often evaluated on live crawls, which are extremely difficult to reproduce, while static corpora typically lack the topical labels required for rigorous benchmarking. We address both limitations with \ds{}, a replayable mirror of \num{621} organization websites where every page is labeled via majority vote across three LLMs. 

Evaluation on this dataset confirms that retrieval models behave as expected for the targeted crawling task: dense-retrieval configurations over labeled anchor texts from external sites consistently outperform breadth-first crawling, whereas traditional classification and lexical matching fail to yield significant gains. Crucially, the relative ranking of these approaches holds across held-out sites, demonstrating that \ds{} serves as a reliable and reproducible benchmark for focused crawling. Beyond the scope of this study, the benchmark enables further research into decision-level cost–effectiveness trade-offs and rich link signals beyond anchor text.